\documentstyle[sprocl]{article}

\def\half{{\textstyle{1\over 2}}}
\def\bkappa   { \mbox{\boldmath ${\kappa}$}   }

\def\qoo{q_{00}}
\def\qol{q_{01}}
\def\qlo{q_{10}}
\def\qll{q_{11}}

\def\loo{\lambda_{00}}
\def\lol{\lambda_{01}}
\def\llo{\lambda_{10}}
\def\lll{\lambda_{11}}

\def\ell{\epsilon_{11}}

\def\lnoo{\ln\epsilon_{00}}
\def\lnol{\ln\epsilon_{01}}
\def\lnlo{\ln\epsilon_{10}}
\def\lnll{\ln\epsilon_{11}}

\input epsf
\begin{document}

\title{ ANALYTICAL SOLUTION FOR MULTIVARIATE STATISTICS IN 
        RANDOM MULTIPLICATIVE CASCADES\footnote{
        Talk presented at CF'98, 8th International Workshop
        on Correlations and Fluctuations, 
        M\'atrah\'aza, Hungary, June 14--21 1998.}
}

\author{H.C.\ Eggers$^*$, M.\ Greiner$^\dag$ and P.\ Lipa$^\ddag$}

\address{         
         $^*$Department of Physics, University of Stellenbosch, 
             7600 Stellenbosch, South Africa\\
         $^\dag$Max-Planck-Institut f\"ur Physik komplexer Systeme,
         N\"othnitzerstr.\ 38, D--01187 Dresden, Germany\\
         $^\ddag$Institut f\"ur Hochenergiephysik, Nikolsdorfergasse 18,
         A--1050 Vienna, Austria
}

\maketitle


\abstracts{
It has long been a puzzle how to solve random multiplicative cascade
structures analytically. We present an analytical solution found
recently in the form of a simple pedagogical example of the general case.
}


This contribution is intended to serve as an introduction-by-example for
prospective students and the casual observer. For more details
and greater rigour, the interested reader should consult 
the published papers.\cite{Gre98a,Gre98b}

What are multiplicative random cascade models (MRCM's), and why are they 
relevant?
Turbulence in fluids must necessarily obey the Navier-Stokes
equation, while the QCD lagrangian in principle describes all high-energy
interactions and final states.
Nevertheless, many experimental results cannot be reproduced
from the underlying theories, even though they are known.
We hence look to {\bf models} to reproduce essential
features of physical data and hence point the way.
This approach has a long and fruitful history.\cite{Fri95a}

{\bf Multiplicative} variables do not obey classical
central limit theorems but rather
exhibit large fluctuations even in the limit of large numbers.
Large fluctuations, unexplainable in terms of additive variables,
are seen in the energy dissipation $\epsilon$ in
fully developed turbulence and in the number and density of pions in
the final states of high-energy hadronic collisions.

The models contain {\bf random variables} because distributions
of relevant quantities differ greatly from event to event 
in high energy physics and correspondingly in different spatial
``snapshots'' or pieces of time series of $\epsilon$.

Hierarchical structures or {\bf cascades} are used because
they can easily be made scale invariant and are
hence well placed to mimick the scale invariance seen
in the Richardson cascades, the original Kolmogorov theory, as
well as some deviations from it.
In high-energy $e^+e^-$, lepton-hadron and hadron-hadron collisions, 
particle production can be described at least in part as cascades
of partons and hadrons.

A typical simple MRCM goes as follows. Start with an initial interval
of unit length, on which a total ``energy dissipation'' $\epsilon$
is measured\footnote{
      The variable $\epsilon$ can correspond to different 
      physical quantities for different systems; for our purposes it is
      important only that the $\epsilon$'s must be positive scalars.}.
Divide the interval into two equal pieces of length $\half$
and split $\epsilon$ randomly into a ``left'' piece $\epsilon_0$
and a ``right'' piece $\epsilon_1$. These are calculated
from the ``parent'' $\epsilon$ by 
\begin{equation}
  \label{mtb}
  \epsilon_0 = q_0 \epsilon\,, \qquad\qquad \epsilon_1 = q_1 \epsilon\,,
\end{equation}
where the multiplicative random variables $q_0$ and $q_1$ are
found by rolling the dice, the probability of obtaining a particular 
value for $q_0$ and $q_1$ being determined by the model's
``splitting function'' $p(q_0, q_1)$. For example, the so-called 
$p$-model has the splitting function
\begin{equation}
  \label{mtc}
  p(q_0,q_1)
    =  \half
       \left[
       \delta ( q_0 - (1+\beta) ) + \delta ( q_0 - (1-\beta) )
       \right] \;
       \delta ( q_0 + q_1 -2) \,,
\end{equation}
where $\beta$ is an adjustable parameter determining how strong the
fluctuations will be. In other words, $q_0$ has an equal
chance of being $1{+}\beta$ or $1{-}\beta$, with $q_1$ automatically
taking the other option.\footnote{
      The delta function $\delta( q_0 + q_1 -2)$ ensures that
      $q_0$ and $q_1$ add up to $2$ because $\epsilon_0$ and 
      $\epsilon_1$ are {\bf energy densities} so that the left and
      right {\bf energies} $E_0 = \epsilon_0\cdot(\half)$ and 
      $E_1 = \epsilon_1 \cdot(\half)$ add up to $E = \epsilon/1$.
      Hence we call this model ``energy-conserving''.}
In the next step, the left ``0'' length interval is again split into ``left''
and ``right'' subintervals, and the energy densities are given by
\begin{equation}
  \label{mtd}
  \epsilon_{00} = \qoo\epsilon_0 = \qoo q_0 \epsilon
  \qquad \mbox{and} \qquad
  \epsilon_{01} = \qol\epsilon_0 = \qol q_0 \epsilon \,,
\end{equation}
where the $q$'s are again determined by throwing dice weighted by 
$p(\qoo, \qol)$. The same kind of thing happens in the ``1''
interval also, so that we have four subintervals with four energy
densities $\epsilon_{00}$, $\epsilon_{01}$, $\epsilon_{10}$
and $\epsilon_{11}$. This can be continued for as long as we like:
after $J$ such cascade steps we end up with $2^J$ intervals
with one $\epsilon$ per interval, characterised by some binary
address.

The correspondence between these models and data is established
by pretending that the {\bf measured} $\epsilon$'s in an experimental
data set result directly from such a cascade (of course this is a
big simplification), and then to {\bf compare correlations} between
$\epsilon$'s from different bins to those given by the model. In this
way, one can see whether such models reproduce the correlation
data.

The problem is that so far it has been impossible to find systematically
analytic expressions for correlations in these models; only
a few special cases were solved. Here, we show in a very simple example
how, if you do everything in the {\bf logarithms} of the $\epsilon$'s,
then you can get analytical answers easily.

Assume that we have a cascade of just two steps, $J=2$, so that we
have four intervals. 
The tree for our little example cascade is shown in Figure 1.
\par\bigskip
\begin{center}
\mbox{ }
\epsfysize=37mm \epsfbox{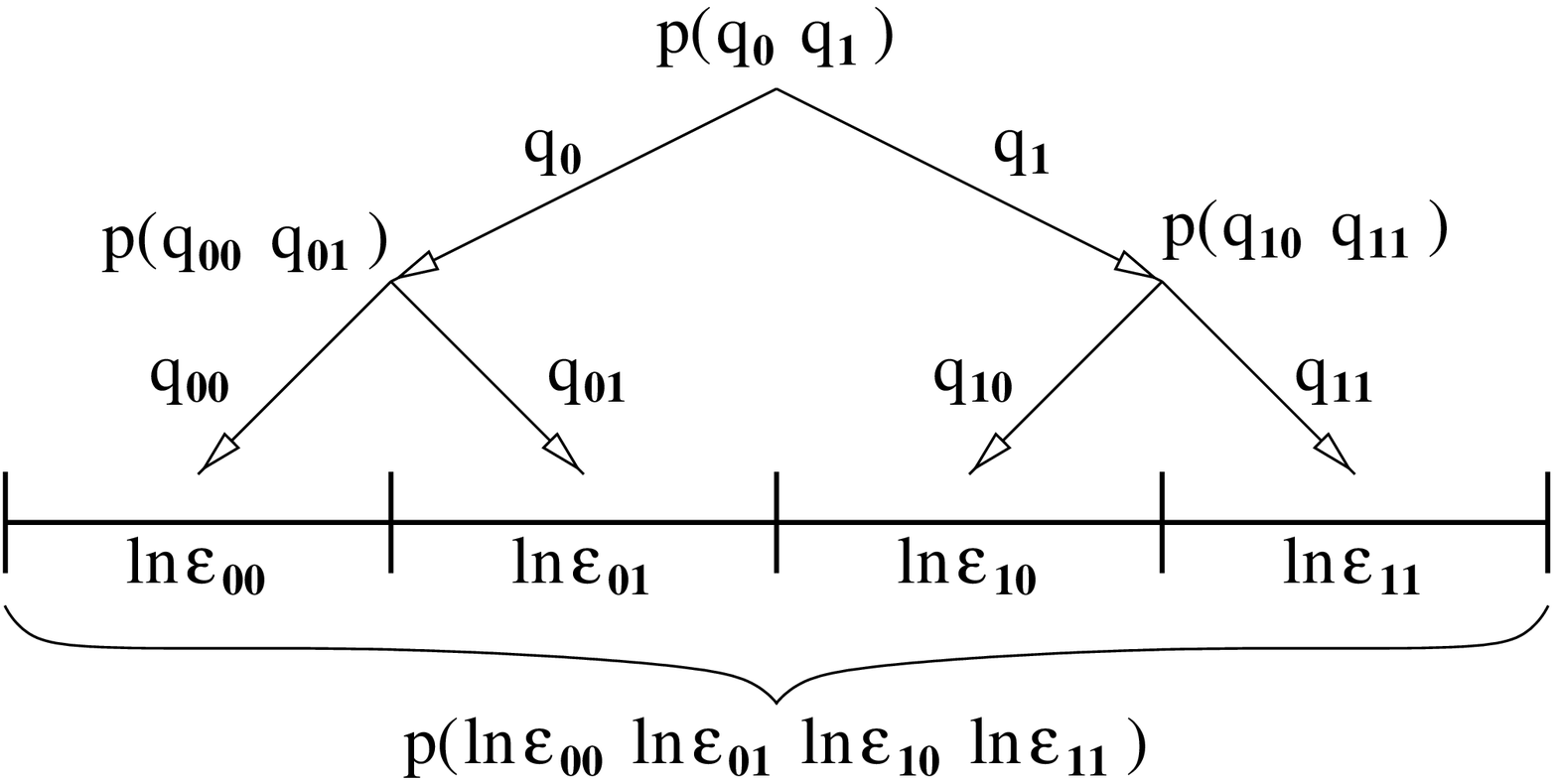}\\
\end{center}
\begin{quote}
{\small Figure 1: Branches, splitting functions $p$,
multipliers $q$ and (logs of)
final energy densities $\ln\epsilon$ for a two-step
multiplicative cascade.}
\end{quote}
The generating function for these four bins,
\begin{equation}
  \label{mte}
  Z_T(\loo,\lol,\llo,\lll) = 
  \left\langle e^{\left( \loo\lnoo + \lol\lnol + \llo\lnlo + \lll\lnll
                    \right)}
  \right\rangle\,,
\end{equation}
contains all possible information on the model, because any
correlation can be found from it (in the form of moments and cumulants)
by taking derivatives of $Z_T$. For example, the ``moment between
bin $00$ and $10$'' is given by
\begin{equation}
  \label{mtf}
  \langle \lnoo \lnlo \rangle 
  =   \rho_{00,10} 
  =   {\partial^2 Z_T \over\partial\loo \partial\llo} 
                 \;\Bigr|_{\loo = \lol = \llo = \lll = 0} \,,
\end{equation}
and the corresponding cumulant is found from
$
  \langle \lnoo \lnlo \rangle 
  - \langle \lnoo \rangle   \langle \lnlo \rangle
  =   C_{00,10} 
  =   ({\partial^2 \ln Z_T / \partial\loo \partial\llo})\,\bigr|_0
$.
Higher orders are calculated in the same way.
Hence, if we know $Z_T$ analytically, all correlations are known
analytically too.

We now show that $Z_T$ factorizes by first writing out the average
$\langle \; \rangle$ in Eq.\ (\ref{mte}) in terms of the joint 
distribution function $p(\lnoo,\lnol,\lnlo,\lnll)$,
\begin{eqnarray}
  \label{mtg}
  Z_T 
  &=& \int d\lnoo\, d\lnol\, d\lnlo\, d\lnll \;
    p(\lnoo,\lnol,\lnlo,\lnll)
    \nonumber\\
  && \times
    e^{\left( \loo\lnoo + \lol\lnol + \llo\lnlo + \lll\lnll \right)} \,.
\end{eqnarray}
Since $\epsilon_{kl} = q_{kl} q_k \epsilon$, the logs are additive,
$\ln\epsilon_{kl} = \ln q_{kl} + \ln q_k$ for all $k,l$. 
(As the initial
energy density is a constant, we can set it to 1 so that $\ln\epsilon = 0$.)
We can hence write $p(\lnoo,\lnol,\lnlo,\lnll)$ in terms of the 
splitting functions by fixing the $\epsilon_{kl}$'s to the appropriate
$q$'s:
\begin{eqnarray}
  \label{mth}
  p(\lnoo,\ldots,\lnll)
  &=& \int dq_0\, dq_1\, d\qoo\, d\qol\, d\qlo\, d\qll\;
  p(q_0,q_1)\, p(\qoo,\qol)\, p(\qlo,\qll)
  \nonumber\\
  &&\quad\times
      \delta[\lnoo {-} \ln\qoo {-} \ln q_0] \;
      \delta[\lnol {-} \ln\qol {-} \ln q_0]
  \nonumber\\
  &&\quad\times    
      \delta[\lnlo {-} \ln\qlo {-} \ln q_1] \;
      \delta[\lnll {-} \ln\qll {-} \ln q_1] \,.
\end{eqnarray}
Inserting this into Eq.\ (\ref{mtg}), we see that the four delta functions
kill the $d\ln\epsilon_{kl}$ integrals, so that we are left with
\begin{eqnarray}
  \label{mti}
  Z_T 
  &=& \int dq_0 \, dq_1 \, p(q_0,q_1)   \,
      \int  d\qoo\, d\qol\, p(\qoo,\qol)\,
      \int  d\qlo\, d\qll\, p(\qlo,\qll)\;
    \nonumber\\
  && \times
    e^{\left[ \loo(\ln\qoo + \ln q_0) 
            + \lol(\ln\qol + \ln q_0)
            + \llo(\ln\qlo + \ln q_1)
            + \lll(\ln\qll + \ln q_1) \right]} \,,
\end{eqnarray}
which can be written in a factorised form:
\begin{eqnarray}
  \label{mtj}
  Z_T(\loo,\lol,\llo,\lll)
  &=&\ \ \
      \int dq_0 \, dq_1 \, p(q_0,q_1)   \,
           e^{\left[ (\loo + \lol)\ln q_0 + (\llo + \lll)\ln q_1
              \right]}
                                                     \nonumber\\
  &&\times
      \int  d\qoo\, d\qol\, p(\qoo,\qol)\,
           e^{\left[ \loo \ln\qoo + \lol \ln\qol \right]}
                                                     \nonumber\\
  &&\times
      \int  d\qlo\, d\qll\, p(\qlo,\qll)\;
           e^{\left[ \llo \ln\qlo + \lll \ln\qll \right]} \,,
\end{eqnarray}
in other words,
\begin{equation}
  \label{mtuj}
  Z_T(\loo,\lol,\llo,\lll)
  =
      Z(\lambda_0, \lambda_1) \;
      Z(\loo,\lol)\;
      Z(\llo,\lll)\,,
\end{equation}
where we have defined $\lambda_0 = \loo + \lol$ and 
$\lambda_1 = \llo + \lll$. Each of the three generating functions 
on the right is the Laplace transform of the corresponding splitting
function; for example
\begin{equation}
  \label{mtk}
  Z(\loo,\lol) = \int  d\qoo\, d\qol\, p(\qoo,\qol)\,
           e^{\left[ \loo \ln\qoo + \lol \ln\qol \right]} \,.
\end{equation}
The importance of Eq.\ (\ref{mtuj}) is that it makes explicit the
statistical independence of the different branchings. This we knew,
of course, because each branching had its own independent splitting
function $p$, but we could not previously make this independence 
explicit in the generating functions.

The independence becomes even more marked when writing (\ref{mtuj}) 
in terms of the {\it cumulant branching generating functions\/}
$Q \equiv \ln Z$: the total cumulant generating function becomes
a simple sum of cumulant branching generating functions (CBGF),
\begin{equation}
  \label{mtl}
  \ln Z_T(\loo,\lol,\llo,\lll) = Q(\lambda_0,\lambda_1)
  + Q(\loo,\lol) + Q(\llo,\lll)\,.
\end{equation}
The above derivation can be generalised to any number of cascade steps:
for $J$ cascade steps leading to $2^J$ bins and $\lambda$'s,
the overall cumulant generating function is the sum of CBGF's, one
for each branching:
\begin{equation}
  \label{mtm}
  \ln Z_T = \sum_{j=1}^J \;\; \sum_{k_1, \ldots, k_{j-1} = 0}^1
 Q(\lambda_{k_1 k_2 \cdots k_{j-1} 0},\lambda_{k_1 k_2 \cdots k_{j-1} 1}) 
       \,,
\end{equation}
where the indices $k_1 k_2 \cdots k_J$ are just the binary addresses of the
different bins as before. Also, just as above we had 
$\lambda_0 = \loo + \lol$ and $\lambda_1 = \llo + \lll$, there
is a ``tree of parameters''
\begin{equation}
  \label{mtn}
  \lambda_{k_1 \cdots k_j} = \sum_{l_{j+1}, \ldots l_J = 0}^1
  \lambda_{k_1 \cdots k_j l_{j+1} \cdots l_J} \,.
\end{equation}
\parbox[t]{60mm}{
The tree structures of (\ref{mtm}) and (\ref{mtn}) are shown in 
Figures 2 and 3. The hierarchy of $Q$'s shows how each one governs
the behaviour of the local branching and how the $Q$'s sum to $\ln Z_T$.
Each $\lambda$ shown in Figure 3 is the sum of all the $\lambda$'s
in the tree below it.

\mbox{\hspace*{0,25in}}
Let us take a concrete example to show what these hierarchies imply.
To find the second cumulant between the bins ``101'' and ``110'',
we note that 
$(\partial \ln Z_T / \partial \lambda_{101})$
contains contributions from the branchings involving $\lambda_1$,
$\lambda_{10}$ and $\lambda_{101}$ because these all ``contain''
the relevant $\lambda_{101}$ through the sum hierarchy of Fig.\ 3.
These contributions are shown by the circled
branching points in Fig.\ 4a. Similarly, the derivative 
$(\partial \ln Z_T / \partial \lambda_{110})$
contains contributions as shown in Fig.\ 4b. Since the lowest circles
in Fig.\ 4a and 4b respectively are completely uncorrelated (their
$Q$'s are independent, after all),
it is thus plausible that only those branching points common to both 
(shown in Fig.\ 4c) will contribute to the second derivative
$(\partial^2 \ln Z_T / \partial \lambda_{101}\, \partial \lambda_{110})$;
indeed this is so. Hence the relevant cumulant can simply be read off as 
}
\hspace*{3mm}
\parbox[t]{60mm}{
\begin{center}
\epsfxsize=54mm \epsfbox{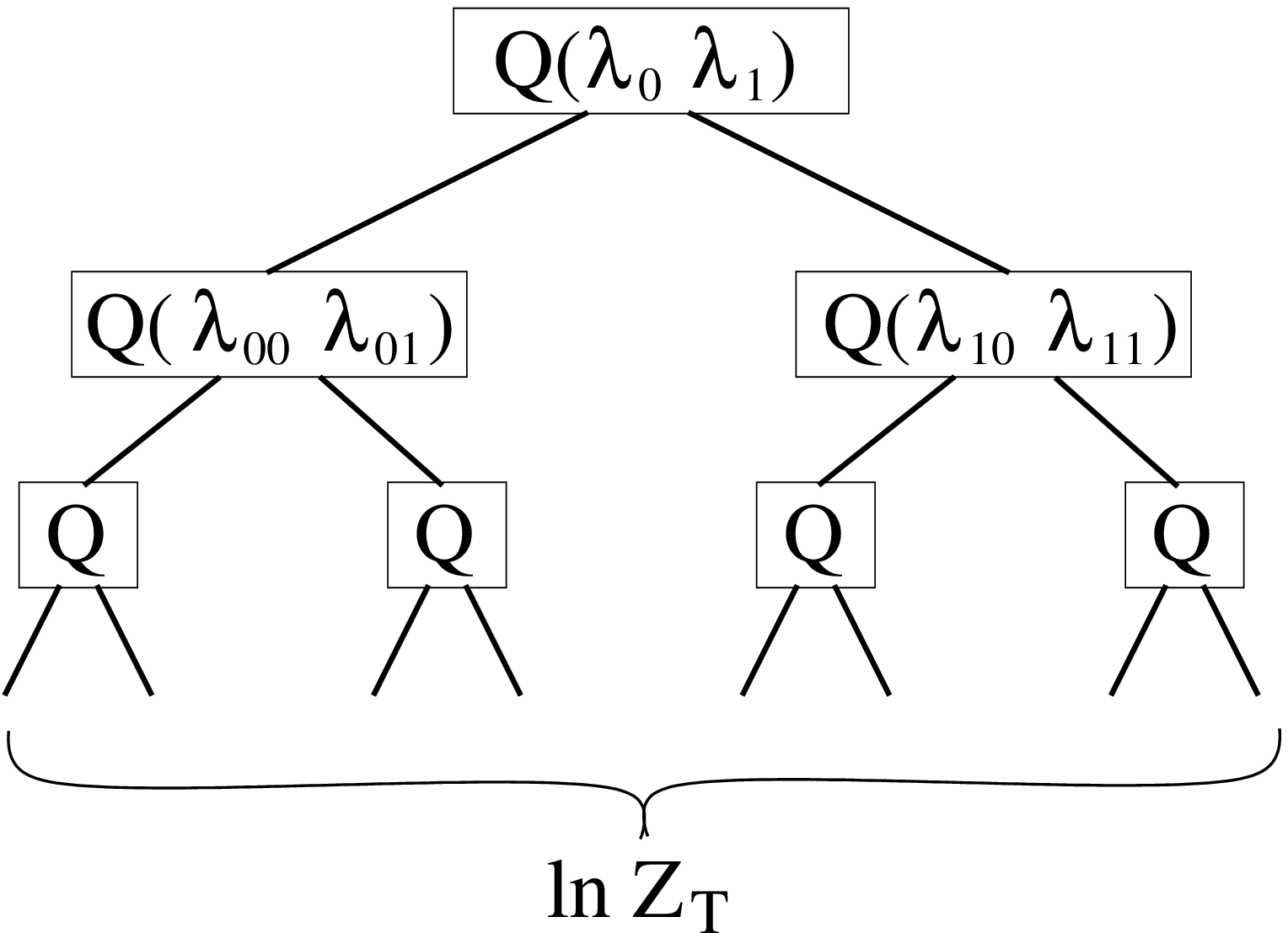}
\end{center}
\vspace*{-7mm}\par
\begin{quote}
{\small Figure 2: Cumulant branching generating functions $Q$,
one at each 
branching point. The overall generating function $\ln Z_T$ is
the sum of all $Q$'s.
}
\end{quote}
\par\vspace*{4mm}\par
\centerline{\epsfxsize=40mm \epsfbox{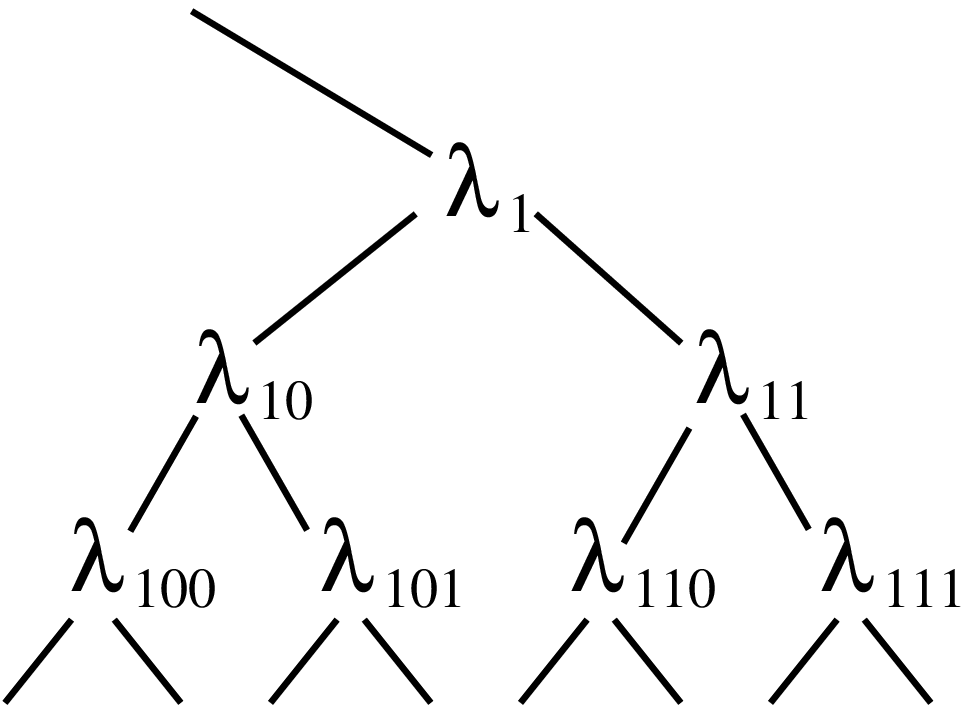}}
\begin{quote}
{\small Figure 3: Each $\lambda$ is the sum of all $\lambda$'s in the tree
below it. }
\end{quote}
}    
\begin{equation}
  \label{mto}
  C_{101,110} 
  = {\partial^2  \ln Z_T \over 
     \partial \lambda_{101}\, \partial \lambda_{110}}\Biggr|_0
  = {\partial^2 Q(\lambda_0,\lambda_1) \over \partial \lambda_1^2 }\Biggr|_0
  + {\partial^2 Q(\lambda_{10},\lambda_{11}) \over 
     \partial \lambda_{10} \, \partial \lambda_{11}}\Biggr|_0 \,.
\end{equation}
\begin{center}
   \parbox[t]{35mm}{
     \begin{center}
       \epsfysize=20mm \epsfbox{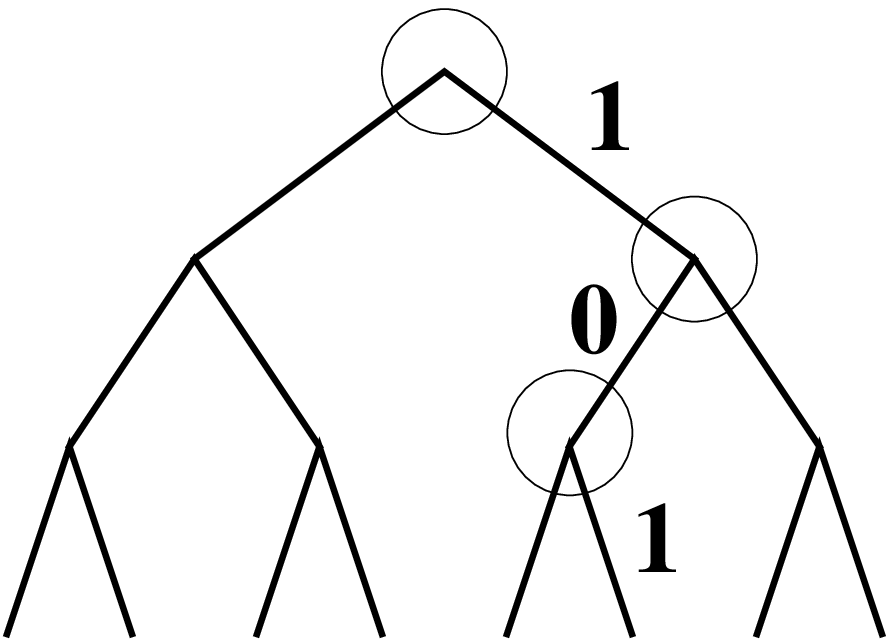}
       (a) \mbox{\hspace*{6mm}}\\
     \end{center}
     }
   \parbox[t]{35mm}{
     \begin{center}
       \epsfysize=20mm \epsfbox{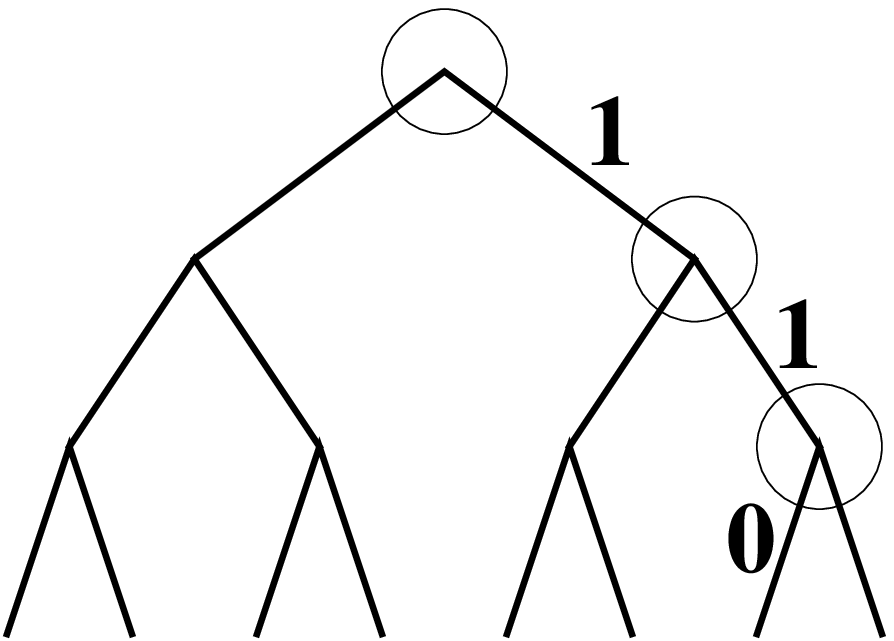}
       (b) \mbox{\hspace*{6mm}} \\
     \end{center}
     }
   \parbox[t]{35mm}{
     \begin{center}
       \epsfysize=20mm \epsfbox{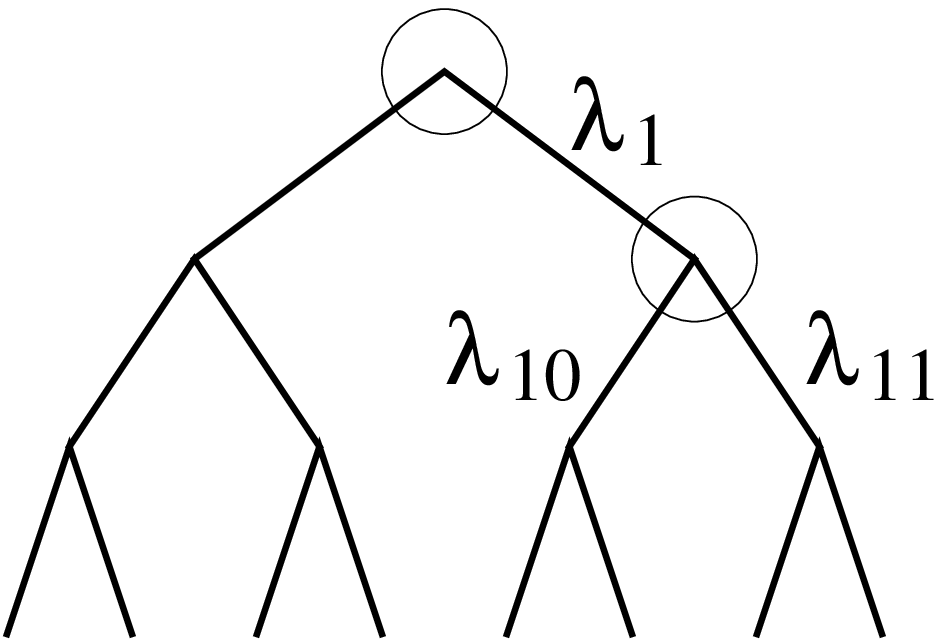}
       (c) \mbox{\hspace*{6mm}}  \\
     \end{center}
     } 
\end{center}
\begin{quote}
{\small Figure 4: Example of calculating the cumulant $C_{101,110}$ from
``same-lineage'' and ``splitting'' cumulant pieces which lie along
the common line of ancestry of the two addresses $101$ and $110$.}
\end{quote}
\noindent
Actually, things become even simpler. Since each $Q$ has only
{\it two\/} parameters, there are two basic kinds of branching cumulants,
namely ``same-lineage'' cumulants, involving only derivatives with respect
to {\it one\/} of the two $\lambda$'s, and ``splitting'' cumulants
involving {\it both\/} the left and right $\lambda$. An arbitrary
multivariate cumulant will then obey the principle of common ancestry:
\par\noindent
\parbox[t]{75mm}{
\mbox{\hspace*{0,25in}}
{\it A cumulant between $n$ bins is the sum of all same-lineage
branching cumulants at branching points which are ancestors of ALL the 
relevant bins. The moment the first splitting in daughters occurs, the 
splitting cumulant of this branching point is added, and then the
process stops.}
In other words, once at least one bin has ``separated'' into a different
lineage, the last common branching point contributes and nothing after
that.

\mbox{\hspace*{0,25in}}
This is expressed elegantly in mathematical form with the help of
the so-called ultametric distance $d_2$, shown in Figure 5, 
which is simply a count of how many generations one must move up 
before a common ancestor is found.
} 
\hspace*{5mm}
\parbox[t]{40mm}{
\begin{center}
       \epsfysize=30mm \epsfbox{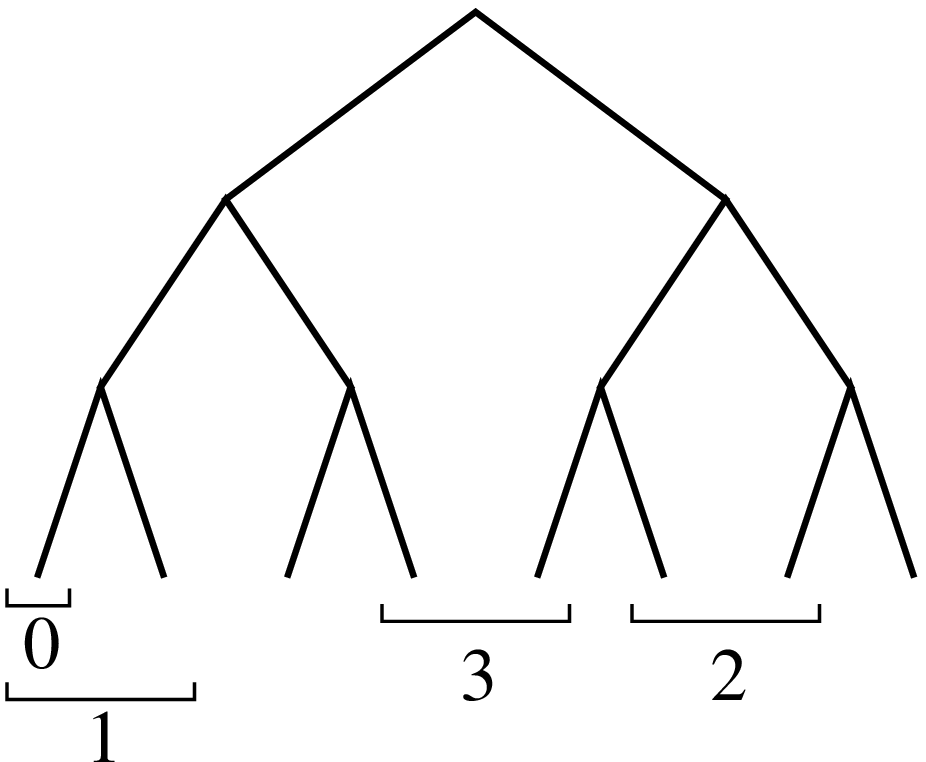}
\end{center}
{\small Figure 5: The ultrametric distance $d_2$ between bins.}
}  
\par\smallskip
The ultrametric distance closely mimicks family terminology,
e.g.\ siblings have $d_2=1$ and cousins $d_2=2$.
For any bin addresses $\bkappa_1$, $\bkappa_2$ and $\bkappa_3$,
the cumulants can be expressed generically as\footnote{
      The assumption made here is that the splitting functions are the 
      same at all branchings. This is not necessary, of course, but the
      formulae would become more complicated.}
\begin{eqnarray}
  \label{mtp}
  C_{\bkappa_1} 
  &=& J\; { \partial\ln Z_T \over \partial \lambda_{\bkappa_1}}\Biggr|_0 \,,
  \\
  \label{mtpt}
  C_{\bkappa_1,\bkappa_2} 
  &=& (J - d_2) {\partial^2 Q \over \lambda_L^2}\Biggr|_0
  + (1 - \delta_{d_2,0})
  {\partial^2 Q \over \partial \lambda_L \partial \lambda_R}\Biggr|_0 \,.
\end{eqnarray}
The second term in (\ref{mtpt}) represents the splitting cumulant
contribution; the $(1 - \delta_{d_2,0})$ makes sure that this is
not counted when $\bkappa_1$ and $\bkappa_2$ are identical. The first
term counts all the same-lineage branching cumulants; the prefactor
$(J - d_2)$ represents the number of common ancestors. 
The same can be done in third order: defining the ultrametric distance
as $d_3 = \max\left[
d_2(\bkappa_1,\bkappa_2), d_2(\bkappa_2,\bkappa_3), \right.\linebreak[4]
       \left. d_2(\bkappa_3,\bkappa_1) \right]$,
the cumulant between any three bins is given by
\begin{equation}
  \label{mtq}
  C_{\bkappa_1,\bkappa_2, \bkappa_3} 
  = (J - d_3) {\partial^3 Q \over \lambda_L^3}\Biggr|_0
  + (1 - \delta_{d_3,0})
  {\partial^3 Q \over \partial^2 \lambda_L \partial \lambda_R}\Biggr|_0 \,,
\end{equation}
with a similar interpretation. Fourth order works similarly.

Let us use the $p$-model as an example again. Insert Eq.\ (\ref{mtc}) 
into the Laplace tranform (\ref{mtk}) and take the logarithm to get
\begin{equation}
  \label{mtr}
  Q(\lambda_0,\lambda_1) =
  \half(\lambda_0 + \lambda_1) \, \ln(1 - \beta^2)
  + \ln\cosh\left[ \half(\lambda_0 - \lambda_1) 
               \ln\left( 1 + \beta \over 1 - \beta \right)
            \right] \,.
\end{equation}
The $\alpha$-model, characterised by
$  p(q_0, q_1) = \textstyle{1\over 4} 
\left[ \delta(q_0 - (1+\beta)) + \delta(q_0 - (1-\beta)) \right]
\linebreak[4]
\times
\left[ \delta(q_1 - (1+\beta)) + \delta(q_1 - (1-\beta)) \right]
$
resembles the $p$-model, except that now the $q_0$
and $q_1$ can take on the values $(1+\beta)$ and $(1-\beta)$ independently.
Its CBGF is found to be
\begin{eqnarray}
Q[\lambda_0, \lambda_1]
&=&
\half (\lambda_0 + \lambda_1) \ln(1 - \beta^2)
 \\
&+& 
\ln   \cosh\left[ \half\lambda_0
        \ln\left( 1 + \beta \over 1 - \beta \right)
       \right]
+
\ln  \cosh\left[ \half \lambda_1
        \ln\left( 1 + \beta \over 1 - \beta \right)
       \right],
\nonumber
\end{eqnarray}
clearly resembling that of the $p$-model, but also clearly different.
The law of common ancestry for these two models is illustrated nicely
in the plot in Figure 6a of the second order cumulant 
$C_{\bkappa_1,\bkappa_2}$ against ultrametric distance $d_2$ between 
bins $\bkappa_1$ and $\bkappa_2$, and in Figure 6b for the fourth order 
$C_{\bkappa_1,\bkappa_2,\bkappa_3,\bkappa_4}$ (note that some indices
are set equal).
Given $J=6$, Eq.\ (\ref{mtpt}) shows that for $d_2=6$ only the 
splitting cumulant contributes: it is negative for the $p$-model
and zero for the $\alpha$-model.\footnote{
     This must naturally be so since $q_0$ and $q_1$ are independent in
     the $\alpha$-model.}
Every time $d_2$ decreases by one, a constant (the same-lineage cumulant)
is added, until $d_2=1$. At $d_2=0$, the (negative) splitting cumulant does
not contribute and so there is a jump in $C_{\bkappa_1,\bkappa_2}$
for the $p$-model. Similar arguments apply in  fourth order.
\par
\noindent
\parbox[t]{60mm}{
\mbox{\hspace*{0,25in}}
We should briefly mention some of the consequences of this analytical
solution. One important connection is that to multifractals, which
we show in Refs.\ \cite{Gre98a,Gre98b} to be equivalent to setting one of
the two variables in $Q(\lambda_0,\lambda_1)$ to zero. 
Hence we contend
that our formalism can ``see more'' than multifractals can: indeed,
the latter cannot distinguish between the $\alpha$- and $p$-models,
while Figure 6 shows that we can. 

\mbox{\hspace*{0,25in}}
In retrospect, it is obvious why the logarithm prescription works: 
by taking the logarithm, the previously multiplicative process becomes
an additive one, which is easily handled and solved.
Previously, conversion from multiplicative
to additive variables could be done only by assuming scaling and
then using scaling exponents,\cite{Pei92a} while the change in variable 
from $\epsilon$ to $\ln\epsilon$ does not rely on scaling 
assumptions at all.
If there is any reason to suspect that multiplicative structures are
at work, it is advisable to work in logarithmic quantities.

} 
\hspace*{2mm}
\parbox[t]{56mm}{
\begin{center}
       \epsfxsize=60mm \epsfbox{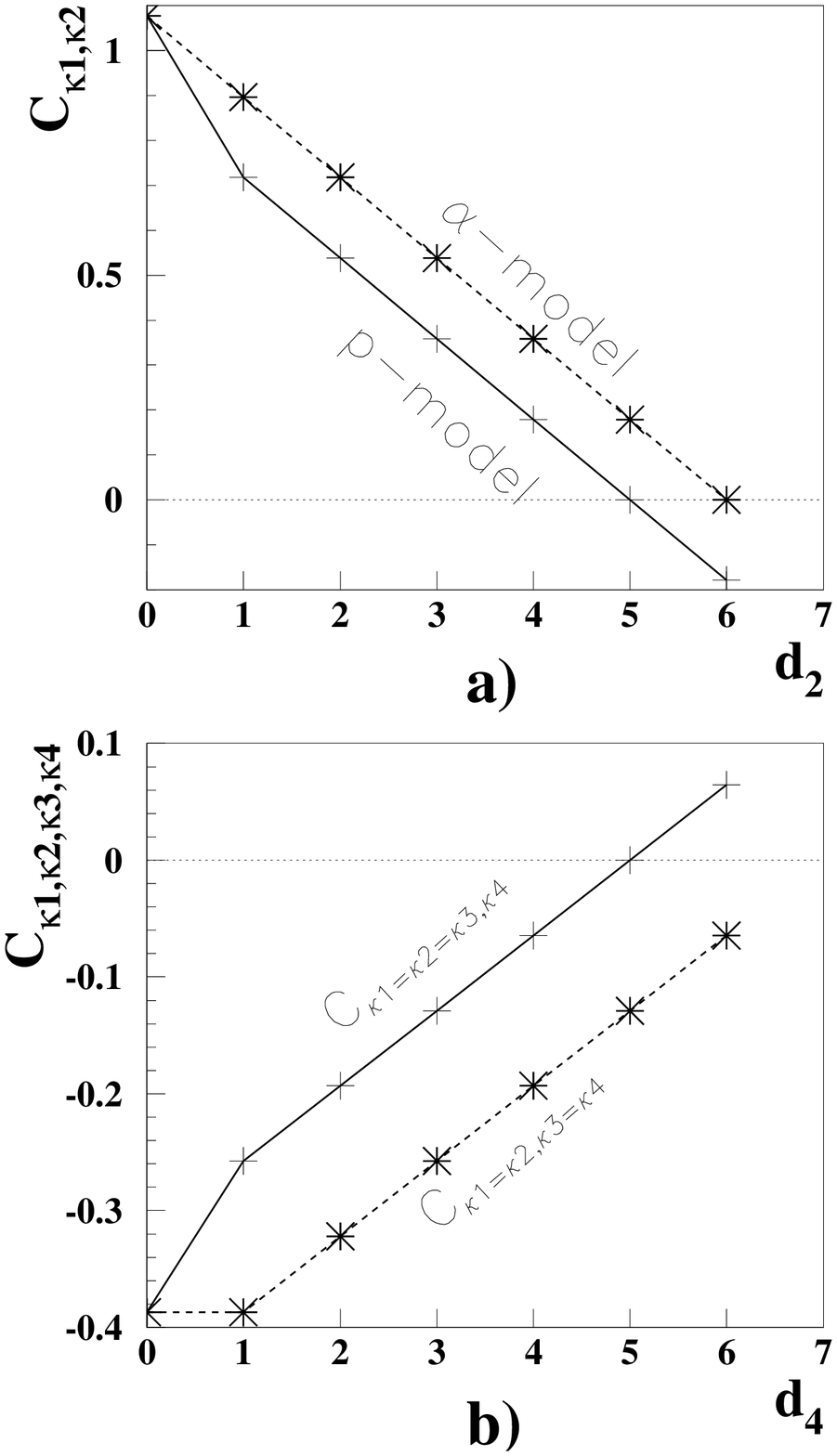}
\end{center}
{\small Figure 6: Cumulants as a function of the 
ul\-tra\-met\-ric dis\-tan\-ces.\hfill\mbox{ }}
}  

\section*{Acknowledgements}
We mourn the loss of Peter Carruthers and dedicate this work to him.
Thanks to the conference organising team for their kind hospitality and
support.
This work was funded in part by the Foundation for Research Development.



\end{document}